\documentclass{article}
\usepackage{ijcai25}

\usepackage{times}
\usepackage{soul}
\usepackage{url}
\usepackage[hidelinks]{hyperref}
\usepackage[utf8]{inputenc}
\usepackage[small]{caption}
\usepackage{graphicx}
\usepackage{amsmath}
\usepackage{amsthm}
\usepackage{booktabs}
\usepackage{algorithm}
\usepackage{algorithmic}
\usepackage[switch]{lineno}
\usepackage{tabularx}

\title{Dual Audio-Centric Modality Coupling for Talking Head Generation}

\author{
    Author Name
    \affiliations
    Affiliation
    \emails
    email@example.com
}

\author{
Ao Fu$^{1,2}$
\and
Ziqi Ni$^{1,2}$\and
Yi Zhou$^{1,2}$\footnote{Corresponding author}\\
\affiliations
$^1$School of Computer Science and Engineering, Southeast University, China\\
$^2$Key Laboratory of New Generation Artificial Intelligence Technology and Its
\\Interdisciplinary Applications, Ministry of Education, China\\
\emails
\{220232248, ziqi\}@seu.edu.cn,
yizhou.szcn@gmail.com
}

\begin{document}
\maketitle

\begin{abstract}
The generation of audio-driven talking head videos is a key challenge in computer vision and graphics, with applications in virtual avatars and digital media. Traditional approaches often struggle with capturing the complex interaction between audio and facial dynamics, leading to lip synchronization and visual quality issues. In this paper, we propose a novel NeRF-based framework, Dual Audio-Centric Modality Coupling (DAMC), which effectively integrates content and dynamic features from audio inputs. By leveraging a dual encoder structure, DAMC captures semantic content through the Content-Aware Encoder and ensures precise visual synchronization through the Dynamic-Sync Encoder. These features are fused using a Cross-Synchronized Fusion Module (CSFM), enhancing content representation and lip synchronization. Extensive experiments show that our method outperforms existing state-of-the-art approaches in key metrics such as lip synchronization accuracy and image quality, demonstrating robust generalization across various audio inputs, including synthetic speech from text-to-speech (TTS) systems. Our results provide a promising solution for high-quality, audio-driven talking head generation and present a scalable approach for creating realistic talking heads.
\end{abstract}

The synthesis of audio-driven talking head videos is a growing research area in computer vision and graphics \cite{musetalk,aniportrait,cvthead,geneface}. Based on audio signals, this task involves generating realistic 3D facial dynamics, including head movements, facial expressions, and lip synchronization. The challenge lies in producing highly accurate facial animations and ensuring consistency and naturalness across various audio conditions. This technology is a foundational component for talking head generation and holds significant promise for applications in virtual environments and digital media. Neural Radiance Fields (NeRF) \cite{nerf}, a powerful framework for 3D scene representation, models scenes as volumetric radiance fields, enabling the synthesis of high-quality images from arbitrary viewpoints. This flexibility makes NeRF a compelling choice for generating talking head videos. 

The application of NeRF to audio-driven talking head synthesis presents unique challenges. Traditional NeRF models are designed for static scenes, which require significant modifications to accommodate dynamic facial movements driven by audio. Recent works have introduced audio signals into NeRF to enable the synthesis of dynamic talking portraits. Despite progress, these methods still face significant limitations in the effective extraction and utilization of audio features. Existing approaches typically rely on pre-trained models from audio-related tasks for feature extraction, such as HuBERT \cite{hubert}, a pre-trained model usually used for automatic speech recognition (ASR), or lightweight audio encoders such as Wav2Lip \cite{wav2lip}. HuBERT, trained via large-scale self-supervised learning, excels at capturing multilevel linguistic and phonetic features, enabling it to capture semantic content from audio effectively. However, it lacks sensitivity to the visual modality. In contrast, the audio encoders used in models like Wav2Lip, which focuses on visual-audio alignment, are highly effective in modeling speech-driven lip movements. However, because of their lightweight design, they struggle to generalize to diverse audio conditions, particularly when faced with variations in speaker styles, tones, and synthetic inputs such as text-to-speech (TTS) outputs. These limitations hinder the realism, generalization, and fidelity of the generated talking head videos, as methods relying on audio pre-trained models struggle to capture fine-grained audio-visual correlations, while end-to-end lip-sync approaches exhibit weaker robustness when handling unseen or synthetic audio signals.

To address the challenges, this paper proposes a novel framework, Dual Audio-centric Modality Coupling (DAMC). This framework innovatively combines two distinct audio feature extractors, each designed to capture different aspects of audio information: content Feature for representing semantic content and dynamic feature for ensuring visual synchronization. By adaptively combining these complementary features, DAMC effectively bridges the gap between audio and visual modalities. Specifically, the framework integrates content features, which encode linguistic and phonetic information from speech, with dynamic features that focus on lip synchronization, enabling precise alignment between facial movements and the corresponding speech dynamics. This approach allows the model to capture subtle variations in language and phonetics while maintaining accurate mouth movements, significantly enhancing the quality and naturalness of the generated talking head videos. In terms of 3D rendering, this work extends the ER-NeRF framework \cite{ernerf} by incorporating the tri-plane mapping technique. This addition improves the detail and texture rendering of facial geometry, ensuring high-quality visual effects in the generated videos. This design not only enhances the temporal consistency of the synthesized content but also boosts the model's generalization capability across a wide range of speakers and synthetic audio inputs, including those generated by text-to-speech (TTS) systems.

Our major contributions are highlighted as follows:
\begin{itemize}
%dual modal
\item We introduce an innovative dual audio-centric modality coupling framework that effectively exploits the semantic information within audio signals while ensuring precise dynamic alignment between facial movements and speech in the visual domain. 

\item By feeding both the Mel-spectrogram and waveform generated during the TTS process into two distinct feature extractors in our framework, we significantly improve the model's ability to synthesize realistic talking head videos driven by TTS-generated audio, ensuring accurate synchronization and natural facial expressions across a variety of synthetic speech conditions.

\item Extensive experiments validate the superiority of our DAMC framework over existing methods. Compared to NeRF-based models, our approach achieves notable improvements in key metrics such as lip synchronization precision, setting a new state-of-the-art performance.

\end{itemize}

\section{Related Work}

\subsection{NeRF-Based Talking Head Generation}
NeRF is a powerful framework for modeling 3D scenes using volumetric rendering functions. Recent advances have extended NeRF to dynamic facial synthesis, particularly for audio-driven talking head generation, enabling models to adapt facial geometry to dynamic audio inputs over time. Early work on 4D facial reconstruction \cite{dnerf} introduced the temporal dimension into NeRF, paving the way for dynamic facial animation.

Building on this foundation, Ad-NeRF \cite{adnerf} conditioned implicit functions on audio features, allowing realistic lip synchronization and speech-driven facial synthesis. DFA-NeRF \cite{dfa} further decoupled lip motion features from input audio and used probabilistic sampling to generate personalized facial animations.

Recent methods like DFRF \cite{dfrf}, RAD-NeRF \cite{rad}, and ER-NeRF \cite{ernerf} focus on improving NeRF's efficiency and rendering quality. DFRF utilizes 2D facial priors to accelerate training, RAD-NeRF decomposes high-dimensional face representations into low-dimensional feature grids for faster inference, and ER-NeRF introduces tri-plane mapping to refine facial geometry and textures. These methods highlight NeRF's growing potential in handling dynamic, audio-driven talking head synthesis.

\begin{figure*}[h]
\centering
\includegraphics[width=\linewidth]{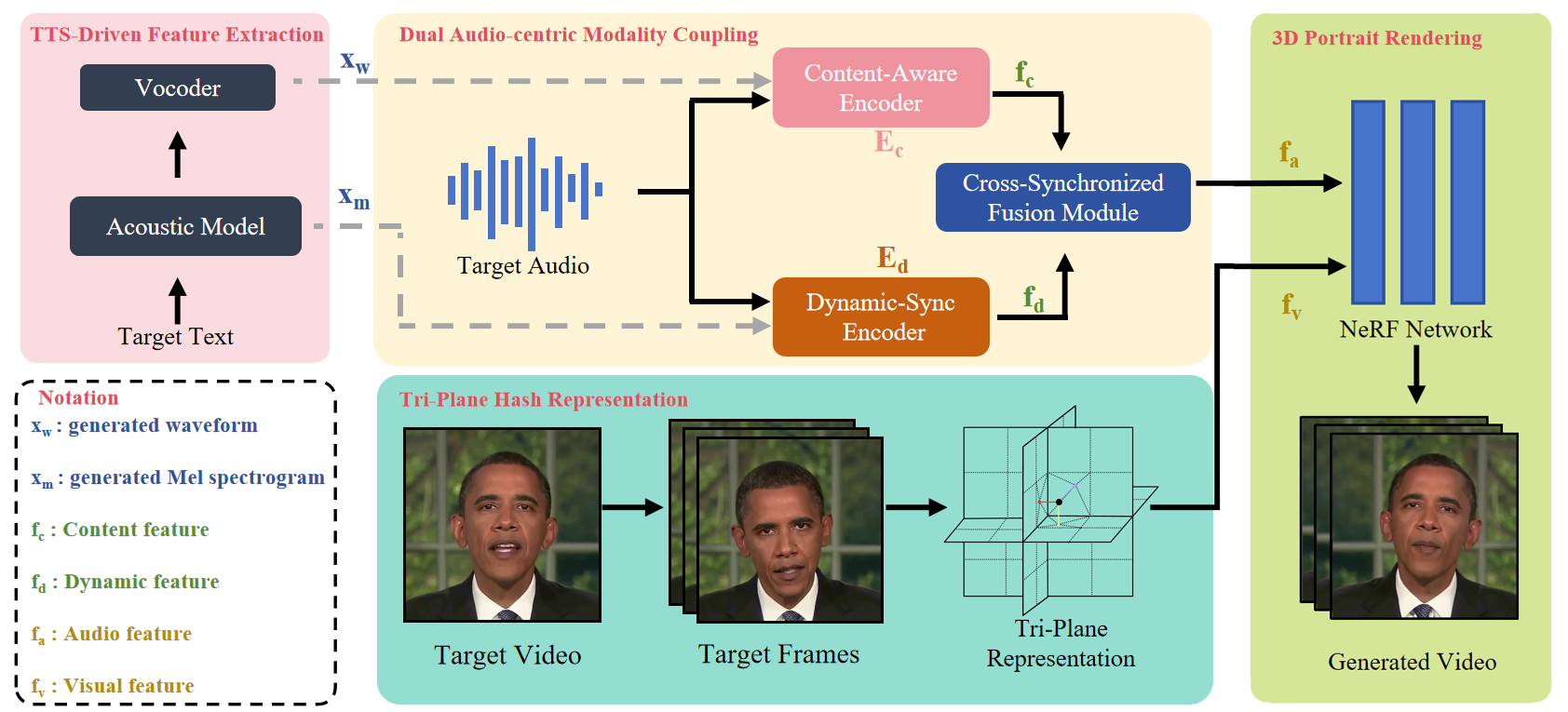}
\caption{\textbf{Overview of the proposed framework.} The framework consists of two modality feature extractors: the Content-Aware Encoder (\( E_c \)) and the Dynamic-Sync Encoder (\( E_d \)), which extract content and dynamic features from the audio, respectively. The extracted features, \( f_c \) and \( f_d \), are fused using the Cross-Synchronized Fusion Module (CSFM). }
\label{fig:1}
\end{figure*}

\subsection{Audio Feature Encoding}
Audio feature encoding is critical in tasks such as ASR, TTS, and talking head generation, as it directly influences the realism and naturalness of the output. Traditional ASR systems employed acoustic features like MFCC and LPCC to capture speech patterns. However, deep learning models such as Wav2Vec \cite{wav2vec}, Wav2Vec 2.0 \cite{wav2vec2}, and HuBERT \cite{hubert} leverage self-supervised learning to extract rich audio representations from large-scale unlabeled datasets. These features excel in speech recognition but primarily focus on semantic content, lacking direct correlations with facial dynamics.

In contrast, talking head generation requires audio features aligned with visual dynamics, particularly lip movements and facial expressions. Recent methods \cite{wav2lip,sadtalker,seeing} align audio and visual modalities using a lip-sync expert, enabling audio features to better capture facial dynamics. By jointly optimizing both modalities, these approaches improve synchronization and naturalness in generated videos.

In this work, we propose the DAMC framework, which couples two distinct audio features—Content Feature, focused on semantic content, and Dynamic Feature, dedicated to visual synchronization. This coupling approach enhances the model's ability to generate highly synchronized, stable, and natural talking head videos, setting a new standard in audio-driven talking head generation tasks.

\section{Methodology}

\subsection{Overview}

%%%%%注意这里能直接把fusion模块称为DAMC吗？
As shown in Figure \ref{fig:1}, we provide a detailed description of the proposed Dual Audio-centric Modality Coupling (DAMC) framework for generating audio-driven talking head videos.
First, in Section \ref{sec:3.2}, we introduce the core components of the DAMC framework, including the Content-Aware Encoder, the Dynamic-Sync Encoder, and the Cross-Synchronized Fusion Module (CSFM), explaining how they collaboratively extract and fuse content and dynamic features from the audio input. In Section \ref{sec:3.3}, we present the TTS-driven feature extraction method, detailing how the integration of the TTS module enhances the model’s adaptability to synthetic speech. Finally, in Section \ref{sec:3.4}, we discuss the NeRF-based 3D rendering technique, outlining how this method is employed to generate high-quality, natural-looking talking head videos. 
%是否有必要写optimization
%Finally, in Section \ref{sec:3.5}, we delve into the optimization and training strategies, focusing on the approaches we adopt during the training process to ensure robust performance across a wide range of audio inputs and facial dynamics.

\subsection{Dual Audio-centric Modality Coupling}
\label{sec:3.2}
\subsubsection{Content-Aware and Dynamic-Sync Encoding}

Formally, we define two modality feature extractors: the \textbf{Content-Aware Encoder} (\( E_c \)) and the \textbf{Dynamic-Sync Encoder} (\( E_d \)). These components are designed to extract complementary information from audio input. Specifically, \( E_c \) focuses on capturing semantic content, leveraging a pre-trained feature extractor from ASR tasks, while \( E_d \) is trained to model synchronization-critical dynamics using the High Definition Talking Face (HDTF) dataset \cite{hdtf}. 

To train \( E_d \), we adopt an alignment-based approach inspired by prior work \cite{lip}, ensuring that audio features are precisely synchronized with corresponding lip motion features. This is achieved by optimizing a model to minimize two key losses: a reconstruction loss, which ensures that the generated video faithfully represents the input audio and visual content, and a lip-sync loss, evaluated by a lip-sync expert discriminator \( D_{\text{lip}} \), which quantifies synchronization quality. Through this process, \( E_d \) learns to extract dynamic features that are tightly coupled with lip movements. 

Next, we use the encoders to extract the content feature \( f_c = E_c(X_a) \) and the dynamic feature \( f_d = E_d(X_a) \) respectively. 

\subsubsection{ Cross-Synchronized Fusion Module}
\begin{figure}[t]
\centering
\includegraphics[width=\linewidth]{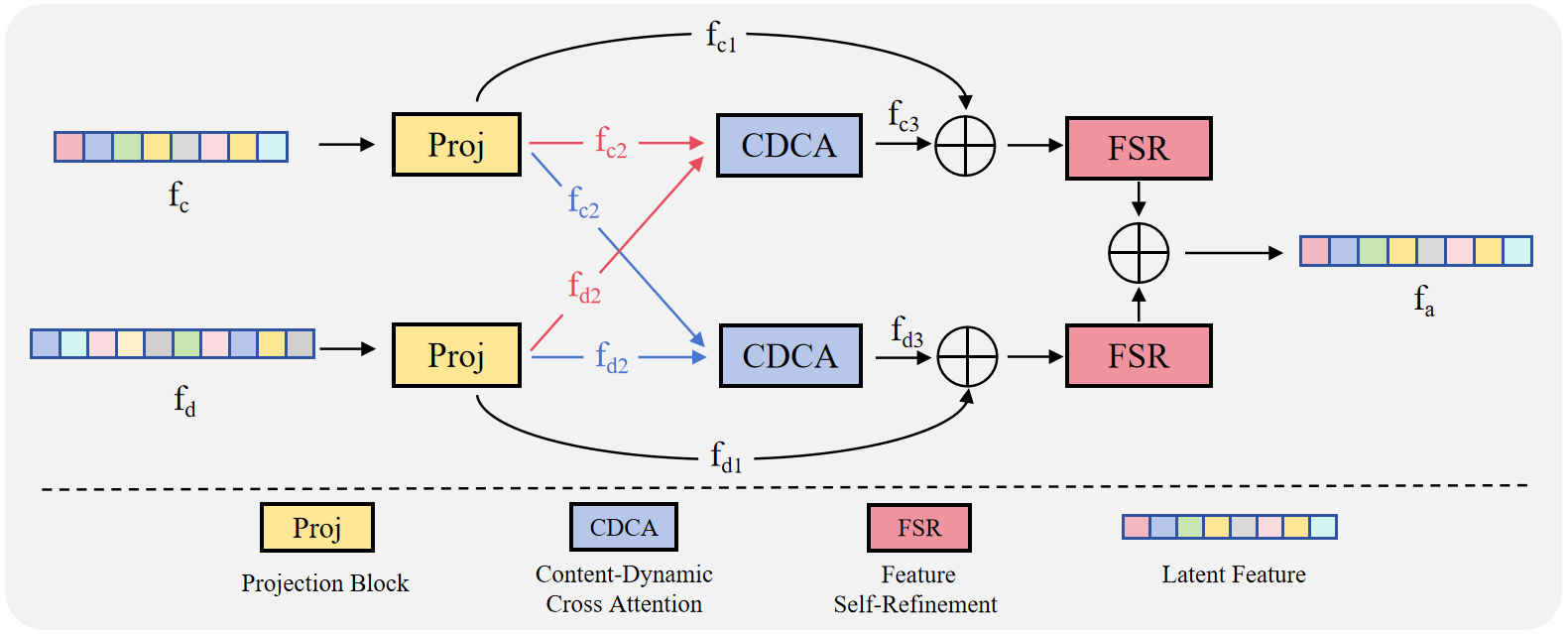}
\caption{\textbf{Structure of the Cross-Synchronized Fusion Module:} The module processes inputs \( f_c \) and \( f_d \) through Projection Block, Content-Dynamic Cross Attention, and Feature Self-Refinement layers to generate the final output \( f_a \).}
\label{fig:2}
\end{figure}

As shown in Fig \ref{fig:2}, to effectively integrate the complementary strengths of the content feature \( f_c \) and the dynamic feature \( f_d \), we propose a \textbf{Cross-Synchronized Fusion Module (CSFM)}. This module leverages Content-Dynamic Cross Attention (CDCA) for inter-modal interaction and Feature Self-Refinement (FSR) to enhance intra-modal consistency. By modeling the interplay between content and dynamic features, the CSFM outputs an audio-driven representation that preserves semantic content while maintaining synchronization dynamics, ensuring a cohesive and precise feature space for our task. 

Initially, the content-aware feature \( f_c \) and dynamic-sync feature \( f_d \) are projected into identical dimensions using modality-specific transformations, producing intermediate representations:

\[
f_{c1} = \text{Proj}_\text{content}(f_c), \quad f_{d1} = \text{Proj}_\text{dynamic}(f_d), 
\]
\noindent while these projection modules are optimized to construct shared representations with minimal computational overhead while retaining the distinct details of each modality.

The cross-modal fusion is achieved via two CDCA modules which facilitate inter-modality feature integration. In the first module, the content-aware feature acts as the query and key, while the dynamic-sync feature serves as the value. This process produces a content-enriched dynamic representation:

\[
f_{c2} = \text{CDCA}(f_{c1}, f_{d1}).
\]

In the second module, roles are reversed, with the dynamic-sync feature as the query and key, and the content-aware feature as the value, resulting in a dynamic-enriched content representation:

\[
f_{d2} = \text{CDCA}(f_{d1}, f_{c1}).
\]

These operations leverage synchronization-aware constraints to establish meaningful dependencies between the modalities, blending the dynamic cues from \( f_d \) into \( f_c \) and vice versa, thus enhancing inter-modal coherence.

Following cross-modal alignment, each feature undergoes the FSR process. This module encapsulates mechanisms inspired by self-attention, extended to emphasize modality-specific consistency while dynamically adapting the features based on the fused information. The refinement process can be expressed as:

\[
f_{c3} = \text{FSR}(f_{c1} + f_{c2}), \quad f_{d3} = \text{FSR}(f_{d1} + f_{d2}).
\]

Here, FSR modules ensure that each modality retains its core attributes while being enhanced by complementary information from the other modality.

The resulting \( f_{c3} \) and \( f_{d3} \) are high-level abstractions that encode audio-centric intra-modal consistency while integrating complementary information across text and visual modalities. These features are subsequently merged into a unified representation \( f_a \), which serves as the final audio-driven embedding. The fusion in CSFM is implicitly guided by the synchronization-aware constraints from the content-aware encoder and dynamic-sync encoder, ensuring robust coupling of speech content and lip motion dynamics. 

This novel fusion mechanism aligns with our framework's overarching objective: bridging content-aware semantic precision and dynamic synchronization fidelity for high-quality audio-driven facial animation tasks. By integrating cross-modal interactions and self-refinement, our DAMC framework ensures a synergistic combination of audio content and motion dynamics.

%To effectively fuse the two input features \( f_c \) and \( f_d \), we design a feature fusion module based on Cross-Attention (CA) and Self-Attention (SA) mechanisms. The process is described as follows:

%First, the input features \( f_c \) and \( f_d \) are processed through their respective multi-layer perceptrons (MLPs) to obtain intermediate representations \( f_{c1} \) and \( f_{d1} \):

%\[
%f_{c1} = \text{MLP}_1(f_c), \quad f_{d1} = \text{MLP}_2(f_d)
%\]

%Next, the inputs \( f_{c1} \) and \( f_{d1} \) are fed into two Cross-Attention (CA) modules for interactive information fusion. The first CA module uses \( f_{c1} \) as the query and key, and \( f_{d1} \) as the value, generating \( f_{c2} \):

%\[
%f_{c2} = \text{CA}(f_{c1}, f_{d1})
%\]

%The second CA module uses \( f_{d1} \) as the query and key, and \( f_{c1} \) as the value, generating \( f_{d2} \):

%\[
%f_{d2} = \text{CA}(f_{d1}, f_{c1})
%\]

%Then, the combined features \( f_{c1} + f_{c2} \) and \( f_{d1} + f_{d2} \) are passed into Self-Attention (SA) modules, which process them to produce updated feature representations \( f_{c3} \) and \( f_{d3} \):

%\[
%f_{c3} = \text{SA}(f_{c1} + f_{c2}), \quad f_{d3} = \text{SA}(f_{d1} + f_{d2})
%\]

%Finally, the output features are obtained by summing \( f_{c3} \) and \( f_{d3} \) to yield the final fused feature \( f_a \):

%\[
%f_a = f_{c3} + f_{d3}
%\]

%This feature fusion mechanism in the DAMC module effectively integrates semantic information from audio content and synchronization information from dynamic features, enhancing the performance of audio-driven facial animation tasks.

\subsection{Inference-Enhanced TTS Module}

\label{sec:3.3}
TTS models play a crucial role in AI systems due to their flexibility and precise control over speech content. We integrate TTS into our framework to enable talking head generation driven by arbitrary text input, allowing for flexible control over speech content. Early TTS frameworks followed a three-stage process: first converting text into linguistic features, then into acoustic features, and finally generating the waveform. Later, TTS systems evolved into a two-stage architecture, where text is transformed into acoustic features through an acoustic model, and then a waveform is generated \cite{ttssurvey}. With the rise of diffusion models \cite{ddpm,ddim}, this two-stage framework not only significantly improved generation quality but also offered greater flexibility.

In our DAMC framework, the input requirements are closely aligned with the evolution of TTS systems. We adopt an ``Acoustic Model-Vocoder" two-stage processing pipeline. During inference, the text is first input into the TTS module, which uses an acoustic model \cite{gradtts,diff,prodiff} to generate mel-spectrogram representations, denoted as \(x_m\). These Mel-spectrograms capture the frequency characteristics of the speech and provide a reliable foundation for subsequent dynamic feature extraction. The Mel-spectrograms are then passed to the dynamic synchronization encoder to extract audio features corresponding to the visual dynamics. Next, the vocoder \cite{wavegrad,diffwave,bddm} decodes the mel-spectrogram into an audio waveform, denoted as \(x_w\), which is resampled and provided as input to the content-aware encoder for extracting content features. This hierarchical processing structure ensures accurate representation of speech content while optimizing the synchronization between audio and facial movements, thereby enabling high-quality speech-driven control for talking head generation.

By integrating TTS-driven inference with the DAMC framework, we not only ensure accurate speech content representation but also optimize the synchronization between audio and facial movements. This combination allows for the generation of talking head driven by arbitrary text inputs while maintaining high-quality outputs. Previous works directly using TTS-generated speech often suffered from unnatural audio and misalignment with the data distribution of the generation model, resulting in poor output quality. In contrast, our approach effectively combines TTS-generated audio features with the DAMC framework, resolving these issues and significantly enhancing the quality of synthesized talking head while maintaining accurate speech expression.

\begin{table*}[t]
    \centering
    \begin{tabularx}{0.9\textwidth}{lXXXXr}
        \toprule
        \textbf{Methods} & 
        \textbf{PSNR $\uparrow$} & 
        \textbf{LPIPS $\downarrow$} & 
        \textbf{FID $\downarrow$} & 
        \textbf{LMD $\downarrow$} &
        \textbf{Sync $\uparrow$} \\
        \midrule

        Wav2Lip \cite{wav2lip}  &    27.123 & 0.068  &  20.911& 5.956 & \textbf{8.932}  \\
        DINet \cite{dinet} & 27.912 & 0.051 &  13.802  & 4.810 & 6.724  \\
        IP-LAP \cite{iplap}       & 30.143 &0.049  & 8.401  &3.933  &4.915  \\
        MuseTalk \cite{musetalk} & 29.530 & 0.053 & 9.417 & 5.014 & 4.291 \\
        AD-NeRF \cite{adnerf}   &27.132& 0.152 & 25.347 & 3.019 & 4.687  \\
        RAD-NeRF \cite{radnerf} &31.852   &  0.069& 9.486 & 2.981 & 4.927 \\
        ER-NeRF \cite{ernerf}       &33.010  & 0.031 &\textbf{4.975}  &3.011  & 5.112 \\
        \midrule
        Ours    &\textbf{33.503}  & \textbf{0.027}& 5.147&\textbf{2.437} & 8.171 \\
        \bottomrule
    \end{tabularx}
    \caption{\textbf{Comparison with State-of-the-Art Methods.} We highlight the \textbf{best} results, with most metrics achieving superior performance.}
    \label{tab:1}
\end{table*}

\begin{table}[t]
    \centering
    \begin{tabularx}{0.45\textwidth}{lXr}
        \toprule
        \textbf{Methods} & &\textbf{Sync $\uparrow$} \\
        \midrule
        Wav2Lip \cite{wav2lip}       & &6.384 \\
        DINet \cite{dinet}           & &4.913 \\
        IP-LAP \cite{iplap}          & &4.037 \\
        MuseTalk \cite{musetalk}     & &3.442 \\ 
        AD-NeRF \cite{adnerf}        & &3.924 \\
        RAD-NeRF \cite{radnerf}      & &4.058 \\
        ER-NeRF \cite{ernerf}        & &4.315 \\
        \midrule
        Ours                         & &\textbf{7.171} \\
        \bottomrule
    \end{tabularx}
    \caption{\textbf{SyncNet Confidence Score of TTS-Driven.} We used randomly generated text as input, driving the model with audio generated by the TTS module. The \textbf{best} results are highlighted.}
    \label{tab:2}
\end{table}

\subsection{NeRF-Based 3D Rendering}
\label{sec:3.4}

High-quality 3D rendering is critical for achieving audio-driven talking head generation. We draw inspiration from tri-plane representation techniques \cite{ernerf} and incorporate a region-aware hash encoding scheme to design an efficient spatial modeling approach, enabling more accurate and efficient dynamic facial rendering.

The tri-plane representation decomposes the 3D space into three orthogonal 2D planes and encodes their features using hash tables, significantly reducing storage and computation overhead. For a given spatial coordinate \( x = (x, y, z) \), the tri-plane representation is defined as:
\[
f_g = H_{XY}(x, y) \oplus H_{YZ}(y, z) \oplus H_{XZ}(x, z),
\]
where \( H_{AB} \) represents the hash encoder for the plane \( AB \), and \( \oplus \) denotes feature concatenation. This decomposition allows the model to construct high-resolution geometric and texture representations at a reduced computational cost while avoiding hash collisions commonly found in traditional methods. Furthermore, the radiance field function used to predict the color \( c \) and density \( \sigma \) is defined as:
\[
F_H(x, d, a; H_3) \to (c, \sigma),
\]
where \( d \) is the viewing direction, \( a \) represents the audio-driven conditional features, and \( H_3 \) denotes the tri-plane hash encoders.

During rendering, the pixel color \( \hat{C}(r) \) along a ray \( r(t) = o + td \) is computed using the volumetric rendering equation:
\[
\hat{C}(r) = \int_{t_n}^{t_f} T(t) \cdot \sigma(r(t)) \cdot c(r(t), d) \, dt,
\]
\noindent where the transmittance \( T(t) \), representing the optical absorption along the ray at depth \( t \), is defined as:
\[
T(t) = \exp\left(-\int_{t_n}^t \sigma(r(s)) \, ds\right).
\]

Here, \( t_n \) and \( t_f \) represent the near and far bounds of the ray's intersection with the 3D scene. By combining volumetric density and color distributions, the rendering process simulates the physical behavior of light propagation to generate the final pixel value.

To optimize this rendering process, we employ a two-stage training strategy \cite{radnerf,ernerf}. In the first stage, coarse-level optimization minimizes the mean squared error (MSE) between the generated and ground truth pixel values for accurate pixel-wise reconstruction. The loss function is given by:
\[
L_\text{coarse} = \sum_{i \in I} \|C(i) - \hat{C}(i)\|^2,
\]
where \( C(i) \) is the ground truth pixel value, \( \hat{C}(i) \) is the generated pixel value, and \( I \) denotes the set of all pixels. In the second stage, fine-level optimization incorporates a perceptual loss (LPIPS) to enhance texture detail and perceptual consistency. The loss function is defined as:
\[
L_\text{fine} = \sum_{i \in P} \|C(i) - \hat{C}(i)\|^2 + \lambda \, \text{LPIPS}(P, \hat{P}),
\]
where \( P \) represents sampled image patches, and \( \lambda \) is a weight for the perceptual loss term.

\section{Experiments}

\subsection{Experimental Setup}

\textbf{Datasets.} We utilize the same video data as previous work \cite{ernerf}, supplemented with high-quality video clips extracted from the HDTF \cite{hdtf} dataset. All videos are 25 frames per second, with most resized to a resolution of \( 512 \times 512 \), except for one video from AD-NeRF \cite{adnerf}, which has a resolution of \( 450 \times 450 \). To train the dynamic-sync encoder, we selected approximately 10 hours of 1080p video from the high-resolution HDTF dataset. These videos were manually filtered to exclude segments with rapid head movements, scenes without human subjects, and other conditions that might degrade data quality. Following the method proposed in \cite{wav2lip}, the curated high-quality video set was used to train the dynamic sync encoder, ensuring the precise capture of the lip dynamics and details of audio-visual synchronization.

\noindent\textbf{Implementation Details:} During the coarse training stage, the model underwent 120,000 iterations, followed by 30,000 iterations in the fine-tuning stage. In each iteration, we sampled \( 256^2 \) rays using a 2D hash encoder (\( L=14, F=1 \)). For the DAMC framework, we experimented with output audio feature dimensions \( f_a \) set to 32, 64, 128, and 256. The best results were achieved when the dimension was set to 64. We use the pre-trained English HuBERT \cite{hubert} model as our content-aware encoder. For TTS-driven audio data generation, we adopt ProDiff \cite{prodiff} as the acoustic model to produce Mel-spectrograms, and WaveGrad \cite{wavegrad} as the vocoder for waveform synthesis. This ensures high-quality and diverse audio inputs for model evaluation. All experiments were conducted on an NVIDIA RTX 4090 GPU, with the total training process taking approximately 3 hours. 

\begin{figure*}[t]
\centering
\includegraphics[width=\linewidth]{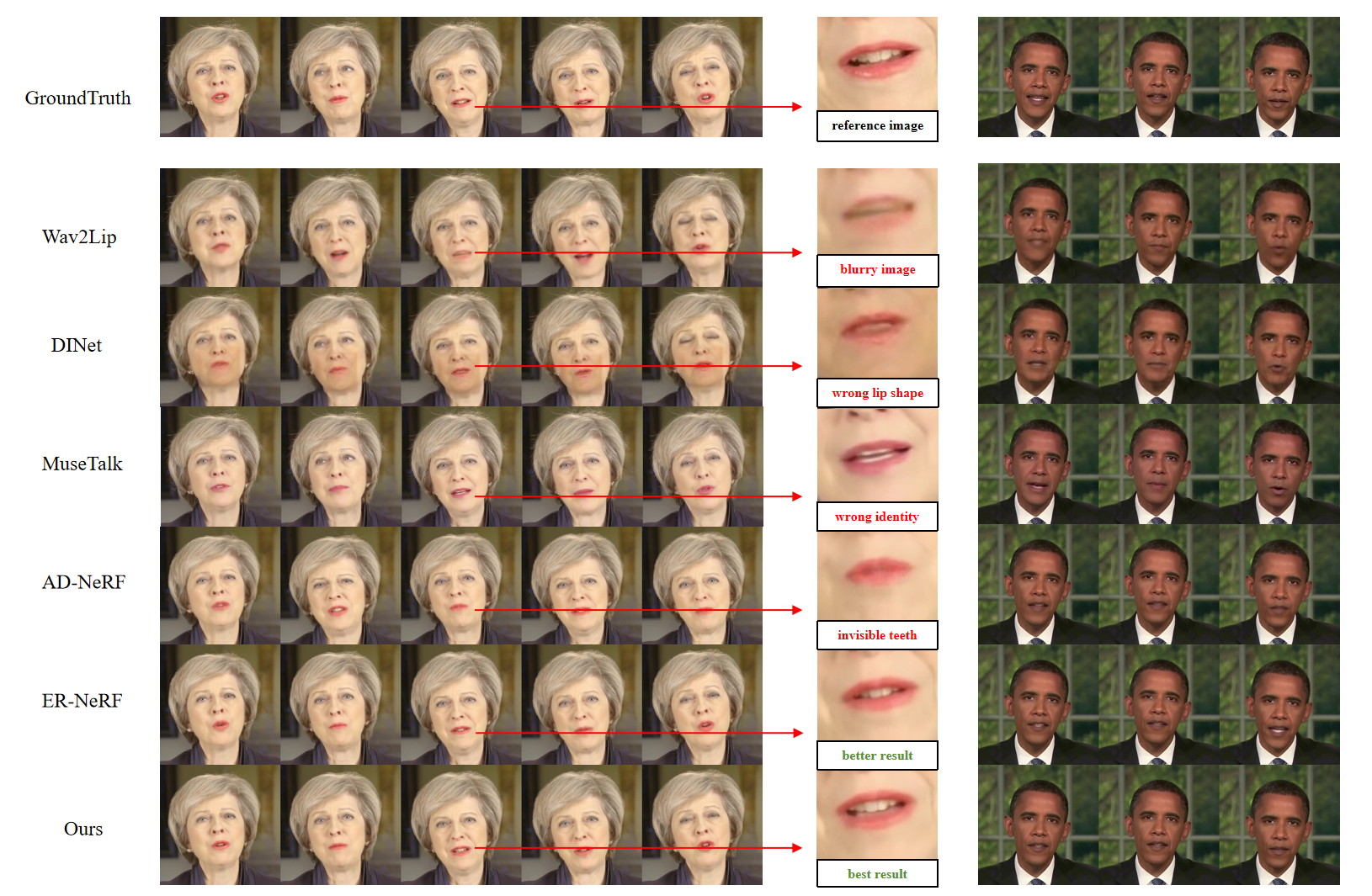}
\caption{\textbf{Qualitative comparison of our proposed method with various state-of-the-art approaches.} The results demonstrate the superiority of our method in terms of visual quality, lip synchronization accuracy, and facial identity preservation.}
\label{fig:3}
\end{figure*}

\subsection{Comparison with State-of-the-Art Methods}
\textbf{Evaluation Metrics.} We use Peak Signal-to-Noise Ratio (\textbf{PSNR}) \cite{PSNR}, Learned Perceptual Image Patch Similarity (Peak Signal-to-Noise Ratio), and Frechet Inception Distance (\textbf{FID}) \cite{fid} as image-level metrics to measure the statistical similarity between the generated video frames and the ground truth at each corresponding timestamp. These metrics provide a comprehensive evaluation of the visual fidelity and perceptual quality of the generated frames. Since the primary goal of our talking head generation task is to ensure facial alignment, smoothness, and naturalness, we include Landmark Distance (\textbf{LMD}) \cite{lmd} as a key metric, which evaluates the geometric accuracy of facial features by computing the deviation of predicted facial landmarks from the ground truth, making it a more meaningful metric for assessing the quality of facial dynamics. We also adopt the SyncNet Confidence Score (\textbf{Sync}) \cite{lip} to evaluate lip synchronization accuracy. This metric quantifies the alignment between audio input and lip movements, serving as a critical measure of temporal and audio-visual consistency.

\noindent\textbf{Evaluation Results.} We compare our proposed DAMC framework with various state-of-the-art (SOTA) models. Common 2D-based models, known for their strong generalization ability, struggle to match the performance of NeRF-based specialized models on specific datasets. 
As shown in Table \ref{tab:1}, our method achieves the highest image quality, with a \textbf{PSNR} of 33.503, surpassing all other methods. In terms of \textbf{LMD}, our method achieves 2.437, which is 18\% lower than the next best result (RAD-NeRF with 2.981), demonstrating precise facial landmark alignment while maintaining high visual quality.

Although Wav2Lip achieves the best \textbf{Sync} score of 8.932, thanks to its use of a pre-trained lip-sync expert, our approach achieves a competitive \textbf{Sync} score of 8.171. Compared to other methods, this represents a 73\% improvement over RAD-NeRF (4.927) and 66\% over ER-NeRF (5.112), while ensuring excellent image quality. These results highlight our method’s ability to balance lip synchronization accuracy with superior image quality.

Furthermore, as shown in Table \ref{tab:2}, our model demonstrates better generalization with TTS-generated speech, achieving superior lip synchronization across diverse audio inputs, including out-of-distribution cases. This underscores our approach’s robustness in handling varied audio sources effectively.

\subsection{Qualitative Evaluation}

To comprehensively evaluate the quality of video generation, we conducted a visual comparison of results from various methods, as shown in Fig \ref{fig:3}. Overall, our approach demonstrates superior performance in terms of image quality, particularly in the fine-grained details of facial features. Compared to methods such as Wav2Lip \cite{wav2lip} and DINet \cite{dinet}, our approach excels in the clarity and precision of lip details. In contrast to MuseTalk \cite{musetalk}, our model effectively preserves identity consistency, avoiding the identity loss issues observed in their results. Furthermore, compared to NeRF-based \cite{adnerf,radnerf,ernerf} methods, our approach achieves significant improvements in lip synchronization while maintaining high video generation quality.

\begin{table}[t]
\centering
\begin{tabularx}{0.45\textwidth}{lXcXc}
\toprule
\textbf{Method} & \textbf{LMD $\downarrow$} & \textbf{Sync $\uparrow$}\\
\midrule
Content Features Only  & 3.331 & 4.417  \\
Dynamic Features Only & 2.976 & 6.624 \\
Concatenation & 6.129 & 3.571 \\
Cross-Attention & 2.911 & 6.697  \\
\midrule
\textbf{Ours} & \textbf{2.718} & \textbf{7.589}  \\
\bottomrule
\end{tabularx}
\caption{\textbf{Ablation Studies about Feature Configurations and Fusion.} Ablation studies highlight the impact of feature design on lip synchronization accuracy (LMD $\downarrow$) and audio-visual synchronization performance (Sync $\uparrow$).}
\label{tab:3}
\end{table}

\begin{figure}[t]
\centering
\includegraphics[width=0.95\linewidth]{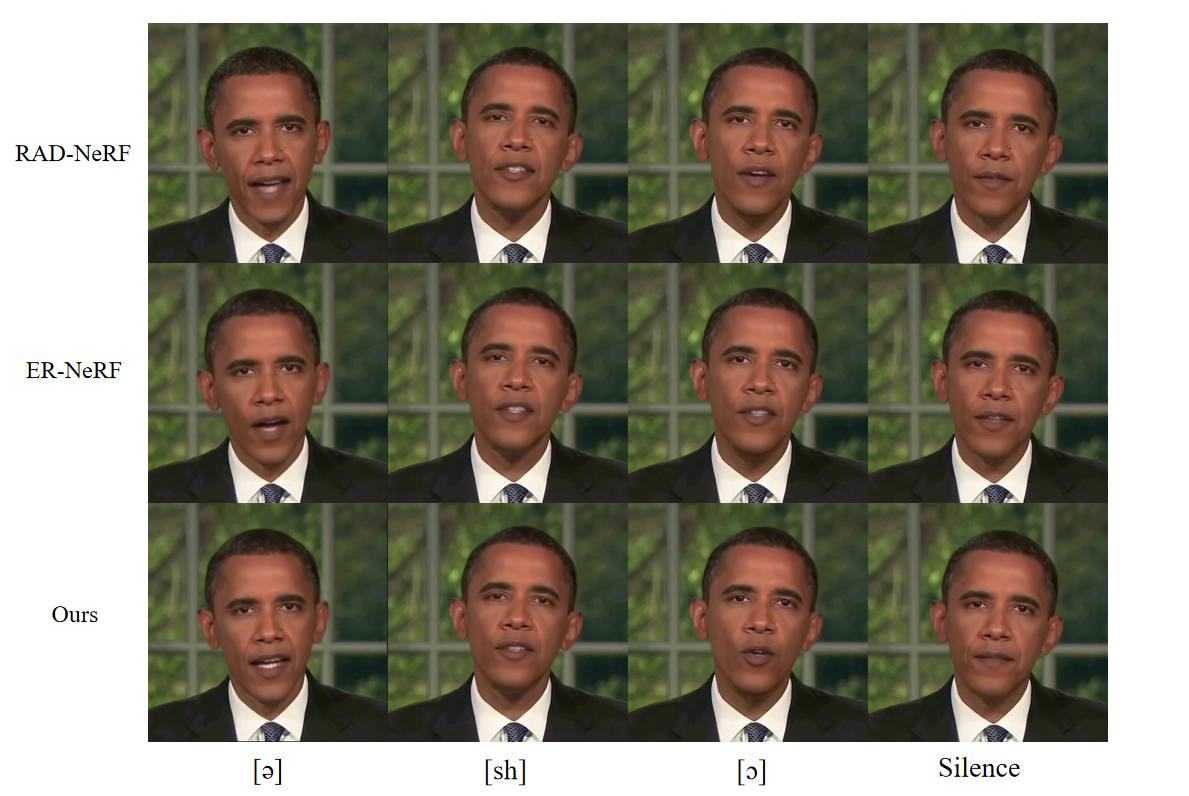}
\caption{\textbf{TTS Driven Video Generation.} Our method maintains relatively excellent lip shapes and image quality even when driven by TTS-generated speech, such as the mouth shapes when producing various phonemes or remaining silent.}
\label{fig:4}
\end{figure}

Notably, in experiments driven by TTS-generated audio, our approach delivers superior lip synchronization, As shown in Fig \ref{fig:4}. Unlike other NeRF-based methods, our model effectively avoids noticeable mismatches between audio and visual outputs, showcasing greater generalization and stability. These findings highlight the advancements of our approach in both visual quality and lip synchronization, establishing a robust foundation for high-quality talking head generation tasks.

\subsection{Ablation Studies}

We conducted comprehensive ablation studies to evaluate the DAMC framework's effectiveness in generating high-quality videos while improving lip synchronization accuracy and generalization to out-of-distribution audio inputs.

The experiments analyzed various feature configurations: using content features alone, dynamic features alone, their direct concatenation, and integration through cross-attention. These were evaluated using two key metrics: \textbf{LMD} for geometric accuracy and \textbf{Sync} for audio-visual synchronization.

As shown in Table \ref{tab:3}, content features alone result in poor synchronization (\textbf{Sync} = 4.417), while dynamic features alone improve it to 6.624 but lack precise geometric alignment (\textbf{LMD} = 2.976). Direct concatenation performs worst (\textbf{LMD} = 6.129, \textbf{Sync} = 3.571), highlighting its ineffectiveness. Cross-attention significantly enhances performance, achieving \textbf{LMD} = 2.911 and \textbf{Sync} = 6.697.

Our method surpasses all configurations, achieving \textbf{Sync} = 7.589 and \textbf{LMD} = 2.718, improving synchronization by 13\% and geometric accuracy by 7\% over cross-attention. These results highlight the importance of our feature design and fusion strategy in balancing semantic representation and synchronization dynamics, achieving state-of-the-art performance.

\begin{table}[t]
\centering
\begin{tabularx}{0.45\textwidth}{lXcXr}
\toprule
\textbf{Method} & &\textbf{Sync $\uparrow$}\\
\midrule
with Pinyin-based method & & 6.107 \\ 
w/o Pinyin-based method & & 5.413 \\
\bottomrule
\end{tabularx}
\caption{\textbf{Impact of Pinyin-based method on lip sync in Mandarin audio.} Experiments showing the effect of using Pinyin-based encoding on lip synchronization accuracy (Sync $\uparrow$).}
\label{tab:4}
\end{table}

\subsection{Case of Mandarin Vocabulary Simplification}

We utilize the audio encoder in the ASR model to build the content-aware encoder, which predicts phoneme probabilities based on a language's vocabulary, meaning feature vector dimensions directly correspond to vocabulary size. Using an English-pretrained encoder for Mandarin audio often results in inaccurate representations, resembling “non-native speaker's clumsy Mandarin.” Fine-tuning for Mandarin is also challenging due to its large vocabulary size, leading to high computational costs and complexity.

To address this, we decompose Mandarin characters into pinyin, reducing the vocabulary size from approximately 3,000 to about 100 phonetic elements (initials and finals). This transformation lowers computational costs and enhances feature expressiveness by encoding audio as phonetic components. Our approach uses an audio encoder from a pre-trained ASR model and maps logits to pinyin with the \texttt{pypinyin} library. As shown in Table \ref{tab:4}, this method improves the fluency and naturalness of Mandarin audio-driven talking head synthesis, delivering smoother and more coherent results.

\section{Conclusion}

In this paper, we proposed the Dual Audio-Centric Modality Coupling (DAMC) framework for audio-driven talking head generation. By effectively combining content and dynamic features through our Cross-Synchronized Fusion Module (CSFM), DAMC significantly improves lip synchronization and visual quality. Experimental results demonstrate that DAMC outperforms existing methods in key performance metrics and shows strong generalization to diverse audio inputs, including synthetic speech. Our framework provides a robust solution for high-quality, audio-driven facial animation and sets the stage for further advancements in virtual avatars and interactive media.

\bibliographystyle{named}
\bibliography{ijcai25}

\begin{thebibliography}{}

\bibitem[\protect\citeauthoryear{Baevski \bgroup \em et al.\egroup }{2020}]{wav2vec2}
Alexei Baevski, Yuhao Zhou, Abdelrahman Mohamed, and Michael Auli.
\newblock wav2vec 2.0: A framework for self-supervised learning of speech representations.
\newblock {\em Advances in neural information processing systems}, 33:12449--12460, 2020.

\bibitem[\protect\citeauthoryear{Chen \bgroup \em et al.\egroup }{2018}]{lmd}
Lele Chen, Zhiheng Li, Ross~K Maddox, Zhiyao Duan, and Chenliang Xu.
\newblock Lip movements generation at a glance.
\newblock In {\em Proceedings of the European conference on computer vision (ECCV)}, pages 520--535, 2018.

\bibitem[\protect\citeauthoryear{Chen \bgroup \em et al.\egroup }{2020}]{wavegrad}
Nanxin Chen, Yu~Zhang, Heiga Zen, Ron~J Weiss, Mohammad Norouzi, and William Chan.
\newblock Wavegrad: Estimating gradients for waveform generation.
\newblock {\em arXiv preprint arXiv:2009.00713}, 2020.

\bibitem[\protect\citeauthoryear{Gafni \bgroup \em et al.\egroup }{2021}]{dnerf}
Guy Gafni, Justus Thies, Michael Zollhofer, and Matthias Nie{\ss}ner.
\newblock Dynamic neural radiance fields for monocular 4d facial avatar reconstruction.
\newblock In {\em Proceedings of the IEEE/CVF Conference on Computer Vision and Pattern Recognition}, pages 8649--8658, 2021.

\bibitem[\protect\citeauthoryear{Guo \bgroup \em et al.\egroup }{}]{rad}
Lidong Guo, Xuefei Ning, Yonggan Fu, Tianchen Zhao, Zhuoliang Kang, Jincheng Yu, Yingyan~Celine Lin, and Yu~Wang.
\newblock Rad-nerf: Ray-decoupled training of neural radiance field.
\newblock In {\em The Thirty-eighth Annual Conference on Neural Information Processing Systems}.

\bibitem[\protect\citeauthoryear{Guo \bgroup \em et al.\egroup }{2021}]{adnerf}
Yudong Guo, Keyu Chen, Sen Liang, Yong-Jin Liu, Hujun Bao, and Juyong Zhang.
\newblock Ad-nerf: Audio driven neural radiance fields for talking head synthesis.
\newblock In {\em Proceedings of the IEEE/CVF international conference on computer vision}, pages 5784--5794, 2021.

\bibitem[\protect\citeauthoryear{Heusel \bgroup \em et al.\egroup }{2017}]{fid}
Martin Heusel, Hubert Ramsauer, Thomas Unterthiner, Bernhard Nessler, and Sepp Hochreiter.
\newblock Gans trained by a two time-scale update rule converge to a local nash equilibrium.
\newblock {\em Advances in neural information processing systems}, 30, 2017.

\bibitem[\protect\citeauthoryear{Ho \bgroup \em et al.\egroup }{2020}]{ddpm}
Jonathan Ho, Ajay Jain, and Pieter Abbeel.
\newblock Denoising diffusion probabilistic models.
\newblock {\em Advances in neural information processing systems}, 33:6840--6851, 2020.

\bibitem[\protect\citeauthoryear{Hore and Ziou}{2010}]{PSNR}
Alain Hore and Djemel Ziou.
\newblock Image quality metrics: Psnr vs. ssim.
\newblock In {\em 2010 20th international conference on pattern recognition}, pages 2366--2369. IEEE, 2010.

\bibitem[\protect\citeauthoryear{Hsu \bgroup \em et al.\egroup }{2021}]{hubert}
Wei-Ning Hsu, Benjamin Bolte, Yao-Hung~Hubert Tsai, Kushal Lakhotia, Ruslan Salakhutdinov, and Abdelrahman Mohamed.
\newblock Hubert: Self-supervised speech representation learning by masked prediction of hidden units.
\newblock {\em IEEE/ACM transactions on audio, speech, and language processing}, 29:3451--3460, 2021.

\bibitem[\protect\citeauthoryear{Huang \bgroup \em et al.\egroup }{2022}]{prodiff}
Rongjie Huang, Zhou Zhao, Huadai Liu, Jinglin Liu, Chenye Cui, and Yi~Ren.
\newblock Prodiff: Progressive fast diffusion model for high-quality text-to-speech.
\newblock In {\em Proceedings of the 30th ACM International Conference on Multimedia}, pages 2595--2605, 2022.

\bibitem[\protect\citeauthoryear{Jeong \bgroup \em et al.\egroup }{2021}]{diff}
Myeonghun Jeong, Hyeongju Kim, Sung~Jun Cheon, Byoung~Jin Choi, and Nam~Soo Kim.
\newblock Diff-tts: A denoising diffusion model for text-to-speech.
\newblock {\em arXiv preprint arXiv:2104.01409}, 2021.

\bibitem[\protect\citeauthoryear{Kong \bgroup \em et al.\egroup }{2020}]{diffwave}
Zhifeng Kong, Wei Ping, Jiaji Huang, Kexin Zhao, and Bryan Catanzaro.
\newblock Diffwave: A versatile diffusion model for audio synthesis.
\newblock {\em arXiv preprint arXiv:2009.09761}, 2020.

\bibitem[\protect\citeauthoryear{Lam \bgroup \em et al.\egroup }{2022}]{bddm}
Max~WY Lam, Jun Wang, Dan Su, and Dong Yu.
\newblock Bddm: Bilateral denoising diffusion models for fast and high-quality speech synthesis.
\newblock {\em arXiv preprint arXiv:2203.13508}, 2022.

\bibitem[\protect\citeauthoryear{Li \bgroup \em et al.\egroup }{2023}]{ernerf}
Jiahe Li, Jiawei Zhang, Xiao Bai, Jun Zhou, and Lin Gu.
\newblock Efficient region-aware neural radiance fields for high-fidelity talking portrait synthesis.
\newblock In {\em Proceedings of the IEEE/CVF International Conference on Computer Vision}, pages 7568--7578, 2023.

\bibitem[\protect\citeauthoryear{Ma \bgroup \em et al.\egroup }{2024}]{cvthead}
Haoyu Ma, Tong Zhang, Shanlin Sun, Xiangyi Yan, Kun Han, and Xiaohui Xie.
\newblock Cvthead: One-shot controllable head avatar with vertex-feature transformer.
\newblock In {\em Proceedings of the IEEE/CVF Winter Conference on Applications of Computer Vision}, pages 6131--6141, 2024.

\bibitem[\protect\citeauthoryear{Mildenhall \bgroup \em et al.\egroup }{2021}]{nerf}
Ben Mildenhall, Pratul~P Srinivasan, Matthew Tancik, Jonathan~T Barron, Ravi Ramamoorthi, and Ren Ng.
\newblock Nerf: Representing scenes as neural radiance fields for view synthesis.
\newblock {\em Communications of the ACM}, 65(1):99--106, 2021.

\bibitem[\protect\citeauthoryear{Popov \bgroup \em et al.\egroup }{2021}]{gradtts}
Vadim Popov, Ivan Vovk, Vladimir Gogoryan, Tasnima Sadekova, and Mikhail Kudinov.
\newblock Grad-tts: A diffusion probabilistic model for text-to-speech.
\newblock In {\em International Conference on Machine Learning}, pages 8599--8608. PMLR, 2021.

\bibitem[\protect\citeauthoryear{Prajwal \bgroup \em et al.\egroup }{2020}]{wav2lip}
KR~Prajwal, Rudrabha Mukhopadhyay, Vinay~P Namboodiri, and CV~Jawahar.
\newblock A lip sync expert is all you need for speech to lip generation in the wild.
\newblock In {\em Proceedings of the 28th ACM international conference on multimedia}, pages 484--492, 2020.

\bibitem[\protect\citeauthoryear{Schneider \bgroup \em et al.\egroup }{2019}]{wav2vec}
Steffen Schneider, Alexei Baevski, Ronan Collobert, and Michael Auli.
\newblock wav2vec: Unsupervised pre-training for speech recognition.
\newblock {\em arXiv preprint arXiv:1904.05862}, 2019.

\bibitem[\protect\citeauthoryear{Shen \bgroup \em et al.\egroup }{2022}]{dfrf}
Shuai Shen, Wanhua Li, Zheng Zhu, Yueqi Duan, Jie Zhou, and Jiwen Lu.
\newblock Learning dynamic facial radiance fields for few-shot talking head synthesis.
\newblock In {\em European conference on computer vision}, pages 666--682. Springer, 2022.

\bibitem[\protect\citeauthoryear{Song \bgroup \em et al.\egroup }{2020}]{ddim}
Jiaming Song, Chenlin Meng, and Stefano Ermon.
\newblock Denoising diffusion implicit models.
\newblock {\em arXiv preprint arXiv:2010.02502}, 2020.

\bibitem[\protect\citeauthoryear{Tang \bgroup \em et al.\egroup }{2022}]{radnerf}
Jiaxiang Tang, Kaisiyuan Wang, Hang Zhou, Xiaokang Chen, Dongliang He, Tianshu Hu, Jingtuo Liu, Gang Zeng, and Jingdong Wang.
\newblock Real-time neural radiance talking portrait synthesis via audio-spatial decomposition.
\newblock {\em arXiv preprint arXiv:2211.12368}, 2022.

\bibitem[\protect\citeauthoryear{Wang \bgroup \em et al.\egroup }{2023}]{seeing}
Jiadong Wang, Xinyuan Qian, Malu Zhang, Robby~T Tan, and Haizhou Li.
\newblock Seeing what you said: Talking face generation guided by a lip reading expert.
\newblock In {\em Proceedings of the IEEE/CVF Conference on Computer Vision and Pattern Recognition}, pages 14653--14662, 2023.

\bibitem[\protect\citeauthoryear{Wei \bgroup \em et al.\egroup }{2024}]{aniportrait}
Huawei Wei, Zejun Yang, and Zhisheng Wang.
\newblock Aniportrait: Audio-driven synthesis of photorealistic portrait animation.
\newblock {\em arXiv preprint arXiv:2403.17694}, 2024.

\bibitem[\protect\citeauthoryear{Yao \bgroup \em et al.\egroup }{2022}]{dfa}
Shunyu Yao, RuiZhe Zhong, Yichao Yan, Guangtao Zhai, and Xiaokang Yang.
\newblock Dfa-nerf: Personalized talking head generation via disentangled face attributes neural rendering.
\newblock {\em arXiv preprint arXiv:2201.00791}, 2022.

\bibitem[\protect\citeauthoryear{Ye \bgroup \em et al.\egroup }{2023}]{geneface}
Zhenhui Ye, Ziyue Jiang, Yi~Ren, Jinglin Liu, Jinzheng He, and Zhou Zhao.
\newblock Geneface: Generalized and high-fidelity audio-driven 3d talking face synthesis.
\newblock {\em arXiv preprint arXiv:2301.13430}, 2023.

\bibitem[\protect\citeauthoryear{Yeung and Werker}{2013}]{lip}
H~Henny Yeung and Janet~F Werker.
\newblock Lip movements affect infants’ audiovisual speech perception.
\newblock {\em Psychological science}, 24(5):603--612, 2013.

\bibitem[\protect\citeauthoryear{Zhang \bgroup \em et al.\egroup }{2021}]{hdtf}
Zhimeng Zhang, Lincheng Li, Yu~Ding, and Changjie Fan.
\newblock Flow-guided one-shot talking face generation with a high-resolution audio-visual dataset.
\newblock In {\em Proceedings of the IEEE/CVF Conference on Computer Vision and Pattern Recognition}, pages 3661--3670, 2021.

\bibitem[\protect\citeauthoryear{Zhang \bgroup \em et al.\egroup }{2023a}]{ttssurvey}
Chenshuang Zhang, Chaoning Zhang, Sheng Zheng, Mengchun Zhang, Maryam Qamar, Sung-Ho Bae, and In~So Kweon.
\newblock A survey on audio diffusion models: Text to speech synthesis and enhancement in generative ai.
\newblock {\em arXiv preprint arXiv:2303.13336}, 2023.

\bibitem[\protect\citeauthoryear{Zhang \bgroup \em et al.\egroup }{2023b}]{sadtalker}
Wenxuan Zhang, Xiaodong Cun, Xuan Wang, Yong Zhang, Xi~Shen, Yu~Guo, Ying Shan, and Fei Wang.
\newblock Sadtalker: Learning realistic 3d motion coefficients for stylized audio-driven single image talking face animation.
\newblock In {\em Proceedings of the IEEE/CVF Conference on Computer Vision and Pattern Recognition}, pages 8652--8661, 2023.

\bibitem[\protect\citeauthoryear{Zhang \bgroup \em et al.\egroup }{2023c}]{dinet}
Zhimeng Zhang, Zhipeng Hu, Wenjin Deng, Changjie Fan, Tangjie Lv, and Yu~Ding.
\newblock Dinet: Deformation inpainting network for realistic face visually dubbing on high resolution video.
\newblock In {\em Proceedings of the AAAI Conference on Artificial Intelligence}, volume~37, pages 3543--3551, 2023.

\bibitem[\protect\citeauthoryear{Zhang \bgroup \em et al.\egroup }{2024}]{musetalk}
Yue Zhang, Minhao Liu, Zhaokang Chen, Bin Wu, Yubin Zeng, Chao Zhan, Yingjie He, Junxin Huang, and Wenjiang Zhou.
\newblock Musetalk: Real-time high quality lip synchronization with latent space inpainting.
\newblock {\em arXiv preprint arXiv:2410.10122}, 2024.

\bibitem[\protect\citeauthoryear{Zhong \bgroup \em et al.\egroup }{2023}]{iplap}
Weizhi Zhong, Chaowei Fang, Yinqi Cai, Pengxu Wei, Gangming Zhao, Liang Lin, and Guanbin Li.
\newblock Identity-preserving talking face generation with landmark and appearance priors.
\newblock In {\em Proceedings of the IEEE/CVF Conference on Computer Vision and Pattern Recognition}, pages 9729--9738, 2023.

\end{thebibliography}

\end{document}